\documentclass[prb,twocolumn,preprintnumbers,superscriptaddress,longbibliography]{revtex4-1}
\usepackage{lineno}
\usepackage{amssymb}
\usepackage{amsmath}
\usepackage{graphicx}
\usepackage[usenames,dvipsnames]{color}
\usepackage[colorlinks=true ,citecolor=blue]{hyperref}

\begin{document}

\title{Time-evolution patterns of electrons in twisted bilayer graphene}

\author{V. Nam Do}
\email{nam.dovan@phenikaa-uni.edu.vn}
\affiliation{Phenikaa Institute of Advanced Study (PIAS), Phenikaa University, Hanoi 10000, Vietnam}
\author{H. Anh Le}
\affiliation{Phenikaa Institute of Advanced Study (PIAS), Phenikaa University, Hanoi 10000, Vietnam}
\author{D. Bercioux}
\affiliation{Donostia International Physics Center (DIPC), Manuel de Lardizbal 4, E-20018 San Sebast\'ian, Spain}
\affiliation{IKERBASQUE, Basque Foundation of Science, 48011 Bilbao, Basque Country, Spain}

\begin{abstract}
We characterise the dynamics of electrons in twisted bilayer graphene by analysing the time-evolution of electron waves in the atomic lattice. 
We perform simulations based on a kernel polynomial technique using Chebyshev polynomials; this method does not requires any diagonalisation of the system Hamiltonian.
 Our simulations reveal that the inter-layer electronic coupling induces the exchange of waves between the two graphene layers. This wave transfer manifests as oscillations of the layer-integrated probability densities as a function of time. For the bilayer case, it also causes a difference in the wavefront dynamics compared to monolayer graphene. The intra-layer spreading of electron waves is irregular and progresses as a two-stage process. The first one characterised by a well-defined wavefront occurs in a short time | a wavefront forms instead during the second stage. The wavefront takes a hexagon-like shape with the vertices developing faster than the edges. Though the detail spreading form of waves depends on initial states, we observe localisation of waves in specific regions of the moir\'e zone. To characterise the electron dynamics, we also analyse the time auto-correlation functions. We show that these quantities shall exhibit the beating modulation when reducing the interlayer coupling.
\end{abstract}

\maketitle

\section{Introduction}
Stacking two-dimensional (2D) materials~\cite{Xu-2013} is a novel method based on the \emph{lego}-principle for creating new van~der~Waals heterostructures with the well-controlled properties.~\cite{Geim-2013} However, accordingly, to the principle of this method, the successive layers are only stacked vertically keeping the same orientation of one to the other. An important step forward comes when allowing a change in the relative orientation of the different stacked layers. The simplest system allowing this new stacking method is twisted bilayer graphene (TBG). This system is composed of two graphene layers stacked within a general manner and has been receiving large consideration lately.~\cite{Rozhkov-2016}  It was predicted that twisting two graphene layers allows a strong tuning its electronic properties.~\cite{Lopes-2009, Santos-2012, Bristritzer-2011,Weckbecker-2016,Koshino-2018} Interestingly, a very narrow isolated energy band around  the charge neutrality level may appear in the spectrum of TBG configurations with tiny twist angles.~\cite{Bristritzer-2011} Recently, Cao \emph{et al.} have experimentally demonstrated that this narrow flat band is responsible to several strongly correlated phases, including an unconventional superconducting and a Mott-like phase.\cite{Cao-2018a, Cao-2018b} Theoretically, it was shown by Zou \emph{et al.} that there are obstructions involving the symmetries of the TBG lattice in constructing effective continuum and tight-binding models to characterise the dynamics of electrons occupying the flat band.~\cite{Zou-2018,Angeli-2018}

Generally, stacking two layered materials may result  in a system of reduced symmetry compared to the two constituent lattices. The atomic configurations of TBG can be characterised by an in-plane vector $\boldsymbol{\tau}$ and a twist angle $\theta$ defining, respectively, the relative shift and rotation between the two graphene lattices. However, it is shown that only the twist angle governs the commensurability of the stacking, regardless of the twisting center.~\cite{Shallcross-2008,Mele-2012,Santos-2012,Zou-2018} In particular, the lattice alignment is commensurate only when the twist angle takes the values given by the formula  $\theta = \text{acos}[(3m^2+3mr+r^2/2)/(3m^2+3mr+r^2)]$, in which $m,r$ are positive coprime integers.\cite{Shallcross-2008, Mele-2012,Santos-2012,Rozhkov-2017,Rode-2017, Zou-2018} When the stacking is commensurate, the translational symmetry of the TBG lattice is preserved, but it usually defines a large unit cell, especially for small twist angles $\theta$.  The electronic calculation for such TBG configurations by brute force diagonalization is therefore extremely expensive in terms of computational resources.\cite{Morell-2010, Trambly-2010,Trambly-2012,Uchida-2014,Lucignano-2019} Furthermore, the electronic calculations based on the time-independent Schr\"odinger/Kohn-Sham equation combined with the Bloch theorem are not applicable for incommensurate configurations because of the loss of the lattice translational invariance. In this work, we show that methods based on the time-dependent Schr\"odinger equation in real space are a powerful alternative to treat the TBG system of arbitrary twist angles.

Following the time-evolution of wave packets in real space is a useful technique to simulate the dynamics of electrons. This method was used for studying the case of monolayer graphene and special TBG configurations. For instance, Rusin and Zawadzki~\cite{Zawadzki-2011} and Maksimova~\emph{et~al.}~\cite{Maksimova-2008} used the kicked Gaussian wave packet to analytically study the different features of the zitterbewegung motion of electrons in various carbon-based structures, including carbon nanotubes. In these works, the wave packets dynamics was governed by an effective  Dirac Hamiltonian, thus the discrete nature of the atomic lattice was not taken into account. M\'ark~\emph{et al.,}~\cite{Mark-2012} however, described the evolution of the kicked Gaussian wave packet in a potential field constructed from an atomistic pseudo-potential model. This approach allows taking into account the distortion of the Dirac cones at high energy, and thus showed the anisotropic dynamics of electrons in the graphene lattice. In the tight-binding framework, Chaves~\emph{et al.}~\cite{Chaves-2010} used the discrete Gaussian form to define a wave packet and showed some quantitative differences in the zitterbewegung motion of electron described by the effective Dirac model and the tight-binding description. Xian~ \emph{et al.}~\cite{Xian-2013} also used the discrete Gaussian wave packet to simulate the transport of electron in a particular commensurate TBG. They showed the existence of six-preferable transport directions along which the wave packets are not broadened, these are along the direction perpendicular to the transport direction.  Particularly, they discussed the behaviour of the layer-integrated probability density in each layer. They  interpreted its behaviour similar to the neutrino-like oscillations where the interlayer coupling plays the role of mixing Dirac fermions in each layer: the two neutrino flavours.

It is well-known that it is the honeycomb structure of graphene as the chiral interlocking of two triangle sub-lattices responsible for all the peculiar properties of graphene and related systems. Accordingly, the electronic properties of graphene can be described by using a formulation in terms of  relativistic fermions.\cite{Neto-2009} This formulation is the same one used by  Schr\"odinger to show the zitterbewegung phenomena as the result of the interference of states at positive and negative energies.~\cite{Schrodinger-1930,Katsnelson-2006}  The two-component spinor structure of the low-energy electron states in graphene is due to the unit cell of the honeycomb lattice constituted by only two carbon atoms. Stacking two graphene sheets gives rise to the diversity of the TBG configurations. It is therefore natural to pose the question of how the manifestation of the atomic lattice structure on the dynamical behaviour of electrons, particularly in the TBG systems with the lack of translational symmetry.

In this work, we address the dynamics of electrons in the real lattice of generic TBG configurations using the tight-binding approach and try to relate it to the lattice symmetries. Though the wave packet method has been successfully used to demonstrate the optical analogy of electrons in graphene,\cite{Jang-2010, Demikhovskii-2010, Rakhimov-2011, Park-2012, Fernandes-2016} its definition depends explicitly on some parameters and therefore not able to provide a full picture of the electronic properties of a system. Accordingly, we will analyse the time-evolution of localised electrons occupying the  $2p_z$ orbitals of carbon atom instead of Gaussian wave packets, whose definition depends on a particular wave vector and an initial position. Within this approach, we can study the changes in the evolution pattern of electron wave functions with respect to the detail of the lattice structure. By artificially tuning the value of the parameters encoding the hybridisation of the $2p_z$ orbitals between two graphene layers, we study the role of the interlayer coupling on the time-evolution of electron states. For studying the time-evolution of a state, we  use the formalism of the time-evolution operator $\hat{U}(t)$, \emph{i.e.}, $|\psi(t)\rangle = \hat{U}(t)|\psi(0)\rangle$; we employ the kernel polynomials method to approximate $\hat{U}(t)$.\cite{Weibe-2006} This method is efficient and useful to work directly in the lattice space of  TBG configurations with arbitrary twist angles. Technically, we use the Chebyshev polynomials of the first kind to approximate the operator $\hat{U}(t)$. Our implementation scheme is efficient because it accounts for the recursive relations of these polynomials, and, as a matter of fact, we are never performing a numerical diagonalization of the system Hamiltonian. Within this method, we can incorporate the details of discrete atomic lattice into the dynamical properties of the $2p_z$ electrons of the TBGs. We shall study the intra-layer development of the $2p_z$ orbitals and the transfer of the probability density from one graphene layer to the other. The local information of the dynamics is studied in the time domain.

The outline of this paper is as follows. In Sec~\ref{method}, we present an empirical tight-binding model which allows characterising the dynamics of the $2p_z$ electrons in different levels of hopping approximation, \emph{i.e.}, the nearest-neighbour (NN), next-nearest-neighbour (NNN) and next-next-nearest-neighbour (NNNN), and we also present the method for the investigation of the time-evolution of the states as well as the calculation for several physical quantities characterising the dynamics of electrons. In Sec.~\ref{results}, we present results for various graphene systems: single-layer graphene, TBG in the AA and AB configuration and finally for various TBG with generic twist angles. Finally, we present conclusions in Sec.~\ref{conclusions}.

\section{Calculation method}\label{method}
In this section, we present the empirical method for defining the tight-binding Hamiltonian for TBG. Subsequently, we present the method for evaluating the time-evolution of a state, and the calculation of the probability density and the density of probability current based on the kernel polynomial method.~\cite{Weibe-2006} Furthermore, we present also a method for evaluating the time auto-correlation function involving the time-evolution of a state, this quantity provides insight on the electronic structure of a system under study. 

\subsection{The empirical tight-binding Hamiltonian}\label{SubA}

The Hamiltonian defining the dynamics of the $2p_z$ electrons  reads:\cite{Do-2018}
%
%
\begin{align}\label{eq2}
\mathcal{H}_\text{TBG}=&\sum_{\nu=1}^2\left[\sum_{i,j}t_{ij}^\nu \hat{c}_{\nu i}^\dagger \hat{c}_{\nu j}+\sum_i V_i^\nu \hat{c}_{\nu i}^\dagger \hat{c}_{\nu i}\right]\nonumber\\
&+\sum_{\nu=1}^2\sum_{ij}t_{ij}^{\nu\bar{\nu}}\hat{c}_{\nu i}^\dagger \hat{c}_{\bar{\nu} j}.
\end{align}
%
%
In this Hamiltonian, the terms in the square bracket define the hopping of the $2p_z$ electron in two graphene monolayers (the index $\nu$ denotes the  layer) with  $t_{ij}^{\nu}$ the intra-layer hopping energies between two lattice nodes $i$ and $j$, and $V_i^\nu$ the onsite energies that is generally introduced to include local spatial effects. The creation and annihilation of an electron at a layer ``$\nu$" and a lattice node ``$i$" is encoded by the operators $\hat{c}_{\nu i}^\dagger$ and $\hat{c}_{\nu i}$, respectively. The hopping of electron between two layers is described by the last term of the Hamiltonian characterised by the hopping parameters $t_{ij}^{\nu\bar{\nu}}$. The notation $\bar{\nu}$ implies that $\bar{\nu}\neq \nu$. The values of the hopping parameters $t^\nu_{ij}$ and $t^{\nu\bar{\nu}}_{ij}$ are obtained via the model:\cite{Moon-2013, Koshino-2015}
%
%
\begin{align}\label{eq3}
t_{ij} = & V_{pp\pi}^0\exp\left(-\frac{R_{ij}-a_\text{cc}}{r_0}\right).\left[1-\left(\frac{\mathbf{R}_{ij}.\mathbf{e}_z}{R_{ij}}\right)^2\right]\nonumber\\
&+V_{pp\sigma}^0\exp\left(-\frac{R_{ij}-d}{r_0}\right).\left(\frac{\mathbf{R}_{ij}.\mathbf{e}_z}{R_{ij}}\right)^2.
\end{align}
%
%
This model for the hopping parameters is constructed through two Slater-Koster parameters $V_{pp\pi} \approx -2.7$ eV and $V_{pp\sigma} \approx 0.48$ eV. These parameters characterise the hybridisation of the nearest-neighbour $2p_z$ orbitals in the intra-layer and inter-layer graphene sheets, respectively. The hopping parameters decay with exponential law as a function of the distance between the lattice nodes $R_{ij} = |\mathbf{R}_{ij}|$; $\mathbf{R}_{ij}$ is the vector connecting two lattice sites $i$ and $j$; $\mathbf{e}_z$ is the unit vector along the $z$-direction perpendicular to the two graphene layers, and $d\approx 0.335$~nm is the distance between two graphene layers. Accordingly, when $i$ and $j$ belong to the same layer, $\mathbf{R}_{ij}$ is perpendicular to $\mathbf{e}_z$ so that we obtain the intra-layer hopping $t_{ij}^\nu = V_{pp\pi}\exp[-(R_{ij}-a_\text{cc})/r_0]$, otherwise we get $t_{ij}^{\nu\bar{\nu}}$. The other parameters are defined as: $r_0\approx 0.184\sqrt{3}a_\text{cc}$ an empirical parameter characterizing the decay of the electron hopping, and $a_\text{cc}\approx 1.42$~\AA\, the distance between two nearest carbon atoms in the graphene lattice. In this work, we are interested in the intrinsic properties of TBG, so we simply set the onsite energies $V_i^\sigma$ to be zero. 

\subsection{The formalisms for the time-evolution of a state}\label{SubB}

Let us start by considering an initial  state $|\psi(0)\rangle$ at the time $t=0$. This state can evolve in time to $|\psi(t)\rangle$ by acting on it with the time-evolution operator $\hat{U}(t) = \exp(-i\hat{H}t/\hbar)$:
%
%
\begin{equation} \label{eq1}
|\psi(t)\rangle = \hat{U}(t)|\psi(0)\rangle = \exp(-i\hat{H}t/\hbar)|\psi(0)\rangle.
\end{equation}
%
%
This equation is  the formal solution of the time-dependent Schr\"odinger equation, where $\hat{H}$ denotes the Hamiltonian operator. We account for the discrete nature of the atomic lattice by describing the system within a tight-binding approximation presented in the Sec.~\ref{SubA}.

In writing the tight-binding Hamiltonian~\eqref{eq2},  we use a localised basis set $\{|j\rangle, j = 1,...,N\}$ to specify the representation. Here, the ket  $|j\rangle$ denotes the $2p_z$ orbital located at the lattice node $j$ and $N$ is the total number of lattice nodes of the whole system. We can express a state $|\psi(t)\rangle$ in this basis set in the following way:
%
%
\begin{equation}\label{eq4}
|\psi(t)\rangle = \sum_{j=1}^N g_j(t)|j\rangle,
\end{equation}
%
%
where $g_j(t)$ determines the probability amplitude of finding electron at the node $j$ at time $t$. The probability density $P_j(t) = |\langle j|\psi(t)\rangle|^2 = |g_j(t)|^2$ is the quantity determining the dynamics of the electron states. The value of $g_j(t)$ is  obtained by solving the time-dependent Schr\"odinger equation or equivalently by performing the calculation for Eq.~\eqref{eq1}.

In this work, we evaluate the time-evolution operator $\hat{U}(t)$ by expanding it in terms of the Chebyshev polynomials of the first kind $Q_m(x) = \cos[m\text{arcos}(x)]$.~\cite{Weibe-2006} As first, we rescale the spectrum of the Hamiltonian $\hat{H}$ to the interval  $[-1,1]$. This scaling is obtained by replacing $\hat{H} = W\hat{h}+E_0$, wherein $W$ is the half of spectrum bandwidth, $E_0$ the central point of the spectrum, and $\hat{h}$ the rescaled Hamiltonian. Practically, we use the power method to estimate $W$. The time-evolution operator is therefore expanded regarding the Chebyshev polynomials as follows:
%
%
\begin{equation}\label{eq5}
\hat{U}(t) = e^{iE_0t/\hbar}\sum_{m=0}^{+\infty}\frac{2}{\delta_{m,0}+1}(-i)^mB_m\left(\frac{Wt}{\hbar}\right)Q_m(\hat{h}),
\end{equation}
%
%
where $B_m$ is the $m$-order Bessel function of the first kind, and $\delta_{m,0}$ is the  Kronecker symbol. We define the so-called Chebyshev vectors $|\phi_m\rangle = Q_m(\hat{h})|\psi(0)\rangle$ which can be calculated using the recursive relation
%
%
\begin{equation}\label{eq6}
|\phi_{m}\rangle = 2\hat{h}|\phi_{m-1}\rangle - |\phi_{m-2}\rangle,
\end{equation}
%
%
with $|\phi_0\rangle = |\psi(0)\rangle$ and $|\phi_1\rangle = \hat{h}|\phi_0\rangle$. Thus, the state $|\psi(t)\rangle$ is formally obtained via:
%
%
\begin{equation}\label{eq7}
|\psi(t)\rangle = e^{iE_0t/\hbar}\sum_{m=0}^{+\infty}\frac{2}{\delta_{m,0}+1}(-i)^mB_m\left(\frac{Wt}{\hbar}\right)|\phi_m\rangle.
\end{equation}
%
%
This equation is exact, but we cannot numerically perform the summation of an infinite series. We therefore approximate $|\psi(t)\rangle$ by a finite series of $M$ terms. Unfortunately, this truncation breaks the preservation of the norm of $|\psi(t)\rangle$. Practically, the number of terms $M$ contributing to the summation in Eq.~\eqref{eq7} is chosen to guarantee the norm conservation of $|\psi(t)\rangle$ in a finite, but sufficiently long, evolution time. For instance, in order to evolve a state in a square TBG sample with 100~nm size for an evolution time of 50~fs, $M$ should be about 1200.\cite{Do-2018}

To define the initial condition for the time-dependent Schr\"odinger equation, one usually assumes the wave function at $t = 0$ of a Gaussian form:~\cite{Nazareno-2007}
%
%
\[\psi_{\mathbf{k}}(\mathbf{r},t=0) = \frac{1}{\sigma\sqrt{\pi}}\exp\left[-\frac{(\mathbf{r-r}_0)^2}{2\sigma^2}\right]\phi_{\mathbf{k}_0}(\mathbf{r}).\]
%
%
In this Gaussian form, $\phi_{\mathbf{k}_0}(\mathbf{r})$ can be simply chosen as a plane wave $\exp(i\mathbf{k_0r})$ or generally as a Bloch function defining a propagating electron state.~\cite{Xian-2013} The Gaussian pre-factor modulates the extension of the function $\phi_{\mathbf{k}_0}(\mathbf{r})$ localised around the position $\mathbf{r}_0$ with a width of $\sigma$. The advantage of this choice is that it allows simulating both the spreading and the moving of the wave centroid. However, the particular behaviour of these phenomena varies concerning $\sigma$ and $\mathbf{k}_0$, two parameters defining a certain initial state. 
 
In this work, we follow a different strategy: we chose a lattice node randomly, then we select the corresponding $2p_z$ orbital to be the initial state. It means that we choose the coefficients $g_j(t=0) = \delta_{ij}e^{i\phi}$, where $\phi$ is a random real number, and thus
%
%
\begin{equation}\label{eq8}
|\psi(t=0)\rangle = \sum_{j=1}^N \delta_{ij}e^{i\phi}|j\rangle = e^{i\phi}|i\rangle.
\end{equation}
%
%
This choice, though does not allow to simulate the displacement of the wave centroid, allows studying the whole energy spectrum of the $2p_z$ electron through the spreading of waves in the various graphene systems. 

To quantify the electron transport, we calculate the expectation value of the probability current operator. In the tight-binding description, the probability current operator reads:~\cite{Do-2018}
%
%
\begin{equation}\label{eq9}
\hat{\mathbf{J}} = \frac{i}{\hbar}\sum_{j,k=1}^N (\mathbf{r}_{j}-\mathbf{r}_{k})t_{jk}\hat{c}^\dagger_j\hat{c}_k.
\end{equation}
%
%
Its expectation value on the state $|\psi(t)\rangle$ is expressed as $\langle\hat{\mathbf{J}}\rangle(t) = \sum_{j=1}^N\mathbf{J}_{j}(t)$ where $\mathbf{J}_{j}(t)$ is interpreted as the density of the probability current:
%
%
\begin{equation}\label{eq10}
\mathbf{J}_{j}(t) = -\frac{1}{\hbar}\sum_{i}(\mathbf{r}_{j}-\mathbf{r}_{i})\text{Im}\left[t_{ij}g_i^*(t)g_j(t)\right].
\end{equation}
%
%

The study of the time-evolution of a state gives us information on the electronic structure of the system. Given an initial state $|\psi(0)\rangle$, the time auto-correlation function $C_\psi(t)$ is defined as the projection of $|\psi(t)\rangle$ on its initial state $|\psi(0)\rangle$:
%
%
\begin{equation}\label{eq11}
C_\psi(t) = \langle \psi(0)|\psi(t)\rangle.
\end{equation}
%
%
In the tight-binding representation with the initial states chosen as a localised  at a particular lattice node $|\psi(0)\rangle = |i\rangle$, the time auto-correlation $C_i(t) = \langle i|\psi(t)\rangle = g_i(t)$, \emph{i.e.}, equal to the local probability amplitude at the node~$i$. Its power spectrum, defined as the Fourier transform of $C_i(t)$, is  the local density of states of electron in the considered system:\cite{Weibe-2006,Do-2018}
%
%
\begin{equation}\label{eq12}
\rho_i(E) = \frac{s}{\pi\hbar\Omega_a}\text{Re}\left[\int_{0}^{+\infty}dt e^{iEt/\hbar}C_i(t)\right]
\end{equation}
%
%
where $\Omega_a$ is the volume assigned for each atom in the lattice and $s = 2$ counts the spin degeneracy. We can obtain the system total density of states from Eq.~\eqref{eq12} by replacing $C_i(t)$ by an ensemble average of $C_i(t)$ over a small set of initial states $|i\rangle$. We implemented this procedure for the first time in Ref.~[\onlinecite{Do-2018}], and results for extremely tiny twist configuration of TBG were in agreement with the approach of continuum models.~\cite{Weckbecker-2016}

\section{Results and discussion}\label{results}
In this section we present results for the three mentioned physical quantities introduced in the previous section: the probability density $P_j(t)$, the density of probability current $\mathbf{J}_j(t)$, and the time auto-correlation function $C_i(t)$, to characterise the dynamics of electrons in  monolayer  and in bilayer graphene systems.

\subsection{Monolayer graphene}
%
%
\begin{figure*}[!t]
\begin{center}
\includegraphics[width = \textwidth]{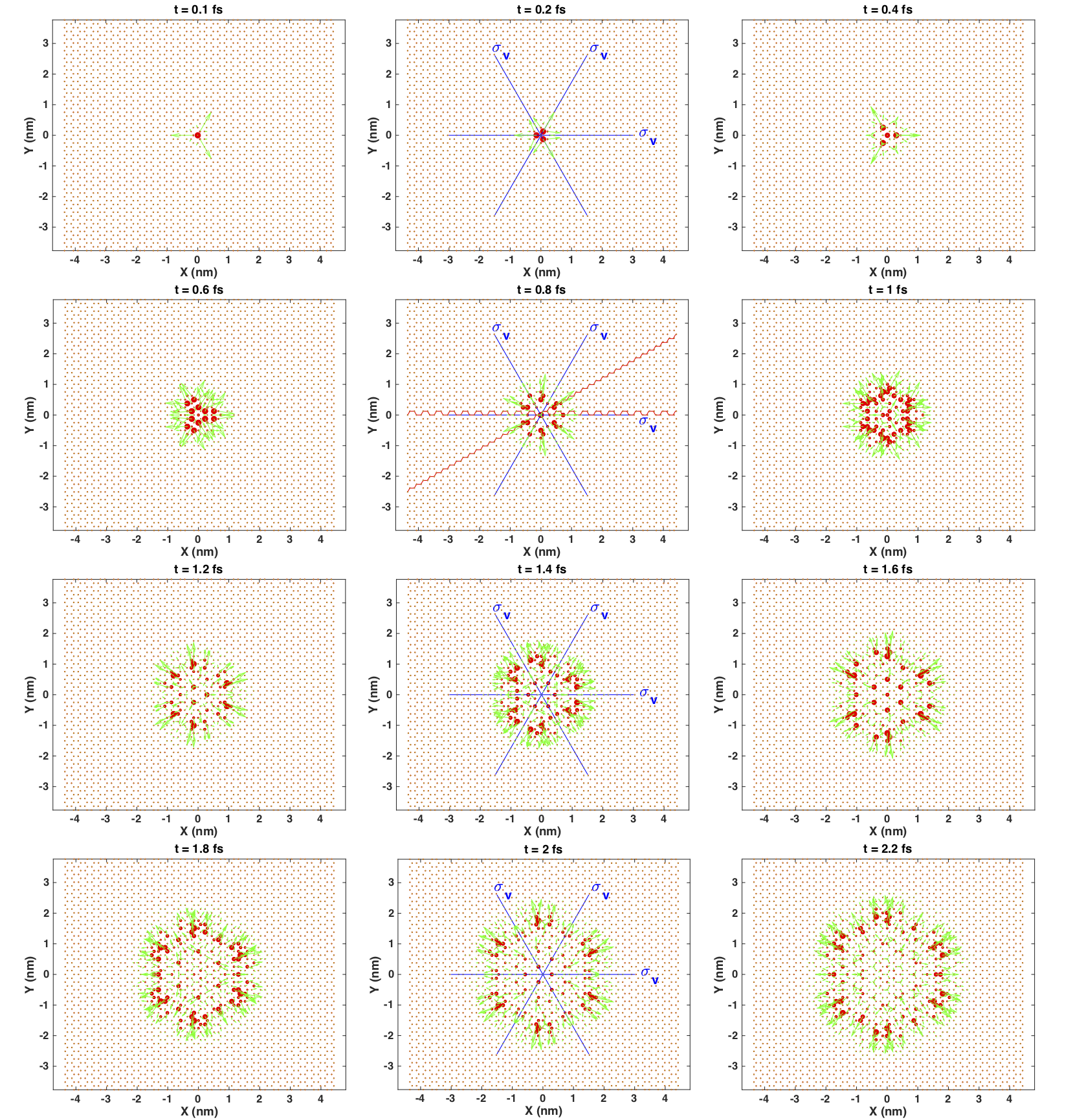}\caption{\label{Fig1}Spreading of the distribution of electron probability densities (in red) in the monolayer graphene taken at several time moments. The arrows denote the vectors of probability current density (in green). The blue lines denote the three mirror-symmetry planes $\sigma_v$ of the hexagonal lattice. We highlight the direction of the armchair- and zigzag-lines in red in the frame with $t=0.8$~fs.}
\end{center}
\end{figure*}
%
%
To better understand the physics of TBG, we shall start analysing the more straightforward case of monolayer graphene. We performed the calculation for the tight-binding Hamiltonians accounting for the NN, NNN, and NNNN hopping terms. As we shall see later, the three models result in different quantitative behaviour for the time auto-correlation function but have the same spreading pattern of the electron wave function; thus for simplicity, we will present results only for the NN case.

We  present in Fig.~\ref{Fig1} the distribution of the probability densities $P_j(t) = |g_j(t)|^2$ and the probability current densities $\mathbf{J}_j(t)$ obtained  for the spreading of a $2p_z$ state initially located at a single lattice node. At each lattice node, we use the solid-circles and the arrows to represent the probability densities and the probability current densities, respectively. The circle radius is proportional to the value of $P_j(t)$, which is normalised at each $t$ to the maximal  value of the set $\{P_j(t)\},\forall j \in\{1,...,N\}$. Similarly, the arrow length is proportional to the length of $\mathbf{J}_j(t)$, which is also scaled appropriately.  The length and the direction of the arrows indicate the tendency that the probability density at a lattice node to transfer to the neighbour lattice nodes. 

The time frames taken at  $t = 0.1$ and $0.2$~fs show that the state firstly spreads to the three nearest neighbours oriented by the angles $\pi$ and $\pm\pi/3$, \emph{i.e.}, along the direction of the armchair lines (highlighted in red in Fig.~\ref{Fig1}, frame with $t=0.8$ fs). The instants $t=0.4$ fs and $0.6$ fs show that the probability current density tends to flow from the central point to the outside along the three directions determined by the angles $0$ and $\pm 2\pi/3$, \emph{i.e.}, also along the direction of the armchair lines. However, at $t = 0.8$~fs the dynamic shows clearly six dominant spreading directions of the probability density, orienting along the directions of the angles $\pm\pi/6, \pm\pi/2$ and $\pm 5\pi/6$, \emph{i.e.}, along the zigzag lines of the honeycomb lattice (c.f. Fig.~\ref{Fig1}, frame with $t=0.8$ fs). For the other time frames, at $ t \in[1,2]$~fs, we find a  continuous spreading of the electron wave function, and we observe the formation of a wavefront with the hexagonal shape. After 2~fs, the wavefront is well established with the corners heading the directions of the zigzag lines.

To quantify the pattern of the wave spreading we directly inspected the distribution of both the probability densities $P_j(t)$ and the probability current densities $\mathbf{J}_j(t)$ on the lattice nodes. We learnt that the distribution of these two quantities obeys the features of the point group $D_{3h}$. These symmetry properties are not identical to those of the graphene lattice, described by the symmorphic space group $p6mm$, and thus the point group $D_{6h}$.~\cite{Kogan-2014} However, we should remember that $D_{3h}$ is a sub-group of the one $D_{6h}$, and is the point group of the lattice node. We then conclude that the spreading pattern of electrons depends on not only the lattice symmetry, but also that of the initial state.

To get quantitative information on the energy spectrum of the $\pi$-bands from the observation of the spreading of a $2p_z$ state, we calculated the time auto-correlation function $C_i(t)$ [c.f.~Eq.~\eqref{eq11}]. The Fourier transform of $C_i(t)$ provides the local density of states at the lattice node $i$.~\cite{Do-2018} In our calculation, we found that the time auto-correlation function, though being a complex function in general, gets purely real when we consider  a model with only the NN hopping. In Fig.~\ref{Fig3} we show the value of $C_i(t)$ obtained by invoking the three hopping models. The result shows the oscillating behaviour of $C_i(t)$ as a function of time with decreasing magnitude. This behaviour implies the declining of the correlation at long evolution times. For the NN hopping approximation, the Hamiltonian has only one parameter $t_\text{cc}=V^0_{pp\pi}$ which sets the energy scale. In this case, we find that $C_i(t)$ is periodic with the oscillation pattern remarked in Fig.~\ref{Fig3}. By changing the value of $V_{pp\pi}^0$ and measuring the corresponding frequency $f$, we verified that the frequency is determined by $f= V_{pp\pi}^0/(2\pi\hbar) = 6.5\times 10^{14}$ Hz. By introducing in the Hamiltonian higher order hopping processes, the behaviour of $C_i(t)$ becomes complex, and we cannot find a clear dominant frequency associated to any of the higher-order hopping terms. Fourier transforming $C_i(t)$ via Eq.~\eqref{eq12}  results in the local density of states which shows the electron-hole symmetry in the NN and NNNN model, but not in the NNN model.\cite{Do-2018}
%
%
\begin{figure}
\begin{center}
\includegraphics[clip = true, trim = 1.7cm 7cm 2.5cm 7.2cm,width = \columnwidth]{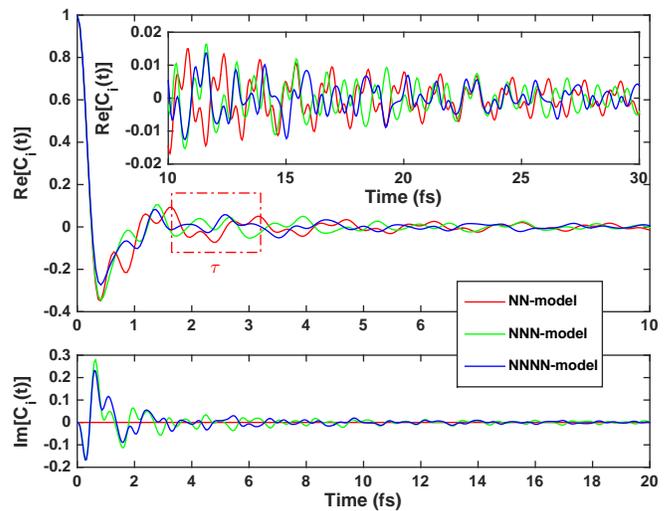}
\caption{\label{Fig2}The real and imaginary part of the auto-correlation function $C_i(t)=\langle i|\psi(t)\rangle$ calculated from three hopping models: the NN in red, the NNN in green and the NNNN in blue. The red rectangle box remarks the typical oscillation pattern predicted by the NN model. The inset shows the continuous variation of the real part of $C_i(t)$ in a longer evolution time.}
\end{center}
\end{figure}
%
%
\subsection{AA- and AB-stacking bilayer graphene}
We will analyse in this section two particular cases of twisted bilayer graphene: the AA- and the AB-stacking. One should notice that we generate the TBG configurations by starting from the AA-stacking configuration and then twisting the two graphene layers about a vertical axis going through a pair of carbon atoms. Accordingly, the AA- and AB-stacking configurations are characterised by a twist angle of~$0^\circ$ and $60^\circ$, respectively.  The point group symmetry of the AA- and AB-bilayer graphene is related to the one of the monolayer. Precisely,  the symmetry of the AA-stacking bilayer graphene is characterised by the symmorphic space group $p6mm$, generated by the lattice translation and the point group $D_{6h}$, whereas the AB-stacking system is characterises by  the symmorphic space group symmetry $p3m1$,~\cite{Latil-2006} generated by the lattice translation and the point group $D_{3d}$.~\cite{Koshino-2010}

\begin{figure}
\begin{center}
\includegraphics[clip = true, trim = 1.8cm 7cm 2.5cm 7.2cm,width = \columnwidth]{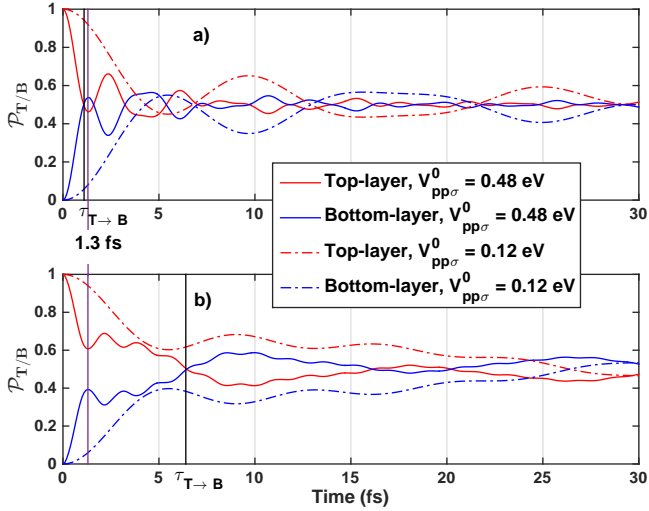}
\caption{\label{Fig3}Neutrino-like oscillation of the layer-integrated probability density $\mathcal{P}_\text{T/B}(t)$ for the (a) AA- and (b) AB-stacking bilayer graphene. Data plotted for two values of the Slater-Koster parameter $V^0_{pp\sigma} = 0.48$ eV (the solid curves) and 0.12 eV (the dot-dashed curves). Only data in the interval of $(0,30)$~fs are displayed to zoom-in the oscillations. The vertical lines highlight the times $\tau_{T\rightarrow B}$ and $t = 1.3$ fs discussed in text.}
\end{center}
\end{figure}

We start by considering the inter-layer transfer of electron wave function: we calculate the layer-integrated probability densities. This quantity is expressed as the summation of the probability density in each layer:
%
%
\begin{align}\label{eq13}
\mathcal{P}_\alpha(t) & = \sum_{j\in (L_\alpha)}P_j(t)  ~~\forall~\alpha\in\{\text{T,B}\}
\end{align}
%
%
In Fig.~\ref{Fig3}, we present  the variation of $\mathcal{P}_\text{T}$ and $\mathcal{P}_\text{B}$ as a function of the time-evolution. The layer-integrated probability density between the two graphene layers presents an oscillatory pattern as a function of time; this behaviour is similar to the phenomenon discussed by Xian~\emph{et al.} as the neutrino-like oscillation.~\cite{Xian-2013} In the case of the AA-stacking configuration, we observe how the wave on the top layer quickly penetrates into the bottom one compared to the AB-stacking configuration. After almost $1.3$~fs, the transfer has reached a maximum of 54\% for increasing then again.  
We notice how different are the oscillatory behaviours for the AA- and AB-stacking configurations from each other, though the distance between the two graphene layers in the two systems  $d = 3.35$~\AA~is the same. The hybridisation of the $2p_z$ orbitals between the two graphene layers is also characterised by the same energy value of $V^0_{pp\sigma}=0.48$ eV. In order to analyse the role of the interlayer coupling on the electron dynamics in the two-layer systems, we investigate the effects of tuning the inter-layer coupling parameter $V^0_{pp\sigma}$ on the layer-integrated probability densities. In Fig.~\ref{Fig3} we present these probabilities obtained by setting $V^0_{pp\sigma} = 0.48$ eV (the solid curves) and 0.12 eV (the dot-dashed curves). 
We observe how the reduction of $V^0_{pp\sigma}$ yields to an increase of the characteristic transfer time, which we define as the evolution time $\tau_{\text{T}\to\text{B}}$ needed to transfer 50\% of the wave from the top layer to the bottom one. Calculation for various values of $V^0_{pp\sigma}$ shows that $\tau_{\text{T}\to\text{B}}\propto 1/V^0_{pp\sigma}$. At very long evolution time, each graphene layer supports about one-half of the initial waves and the layer-interchange transfer becomes almost stationary with very weak oscillations in time. It is worthy to note that, for the AB-stacking configuration, we distinguished two cases of the initial state $|i\rangle$, one at the A-atom on top of the B-atom in the bottom layer, and the other at the B-atom on the center of the hexagonal ring underneath.  We found that the layer-integrated probability densities in the two cases are the same, but the in-plane wave spreading patterns are different as discussed in next paragraphs.
%
%
%
%
%
%
\begin{figure*}
\begin{center}
\includegraphics[width = \textwidth]{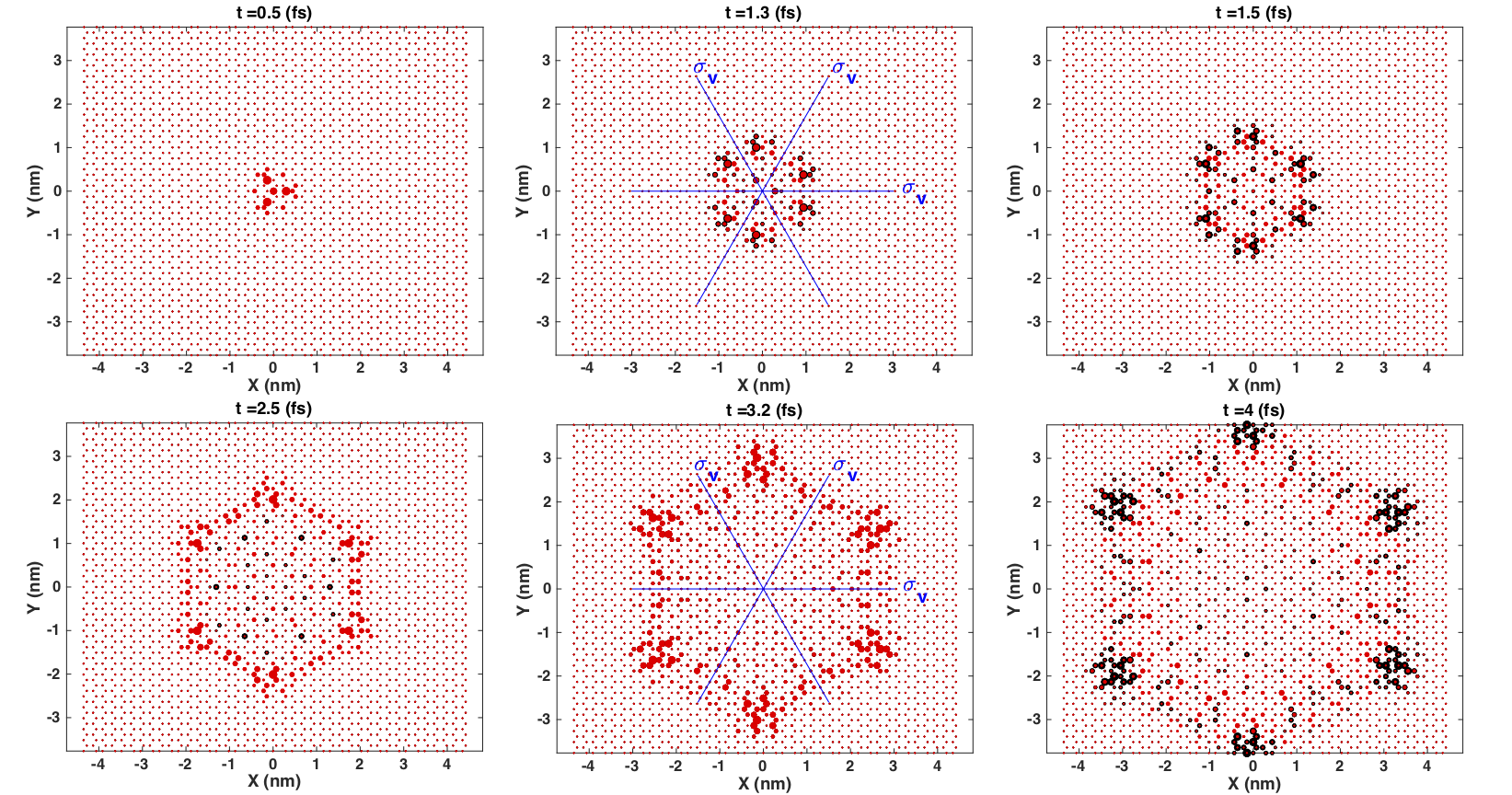}
\caption{\label{Fig4}Spreading of the electron probability density for the AA-stacking configuration taken at several time moments. Lattice nodes in red/black belong to the top/bottom graphene layer. The blue lines in the frames at $t=1.3$ fs and $t = 3.2$ fs denote the three symmetrical mirrors $\sigma_v$ of the lattice.}
\end{center}
\end{figure*}
%
%
%
%
\begin{figure*}
\begin{center}
\includegraphics[width = \textwidth]{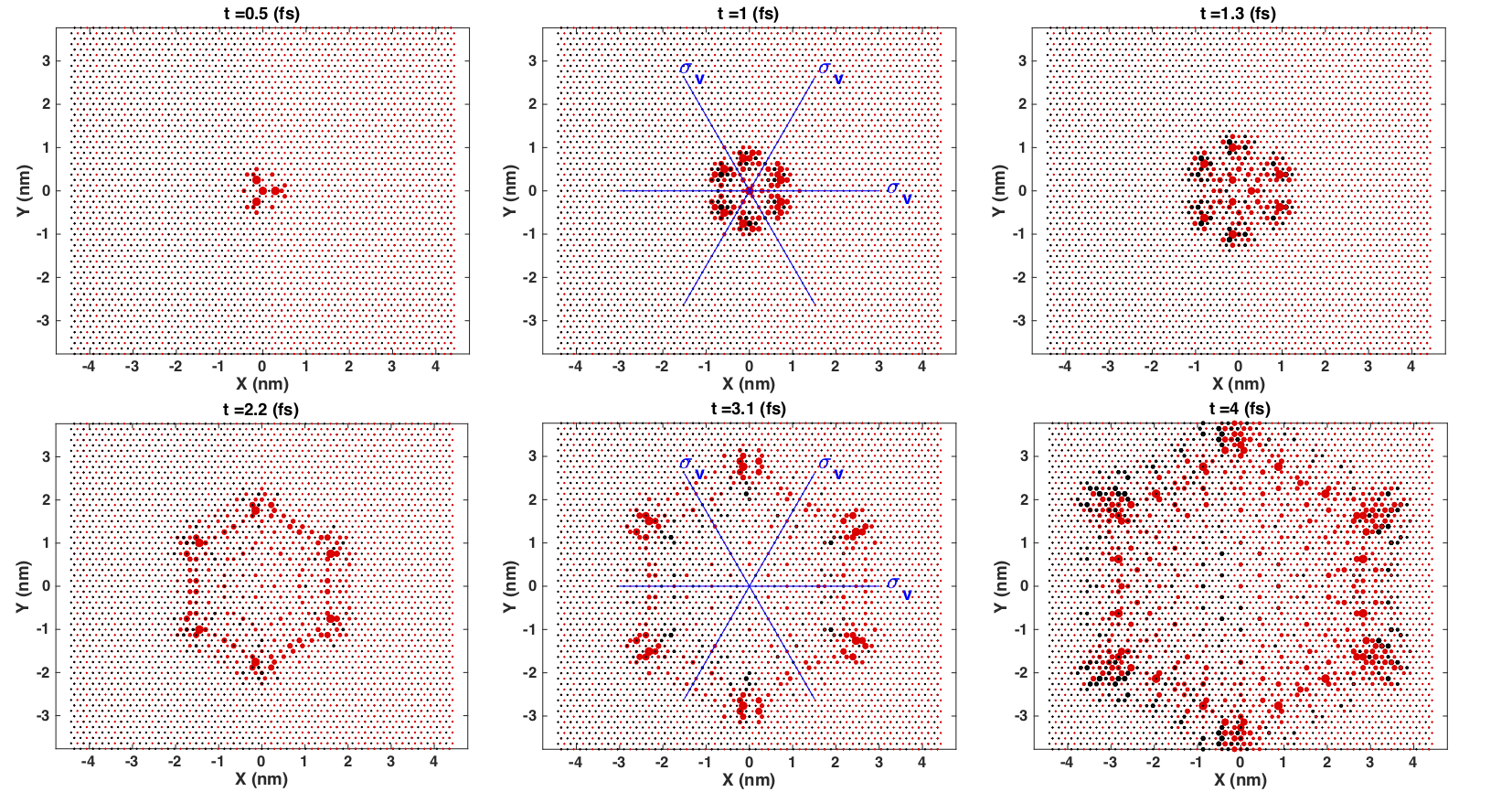}
\caption{\label{Fig5}Spreading of the electron probability density for the AB-stacking configuration taken at several time moments. The initial state $|i\rangle$ is set at the position of an A-atom of the top graphene layer which is on top of a B-atom of the bottom layer. Lattice nodes in red/black belong to the top/bottom graphene layer. The blue lines in the frames at $t = 1$ fs and $t=3.1$ fs denote the three symmetrical mirrors $\sigma_v$ of the lattice.}
\end{center}
\end{figure*}
%
%
%
%
\begin{figure}
\begin{center}
\includegraphics[clip = true, trim = 1cm 7cm 2cm 7cm,width = \columnwidth]{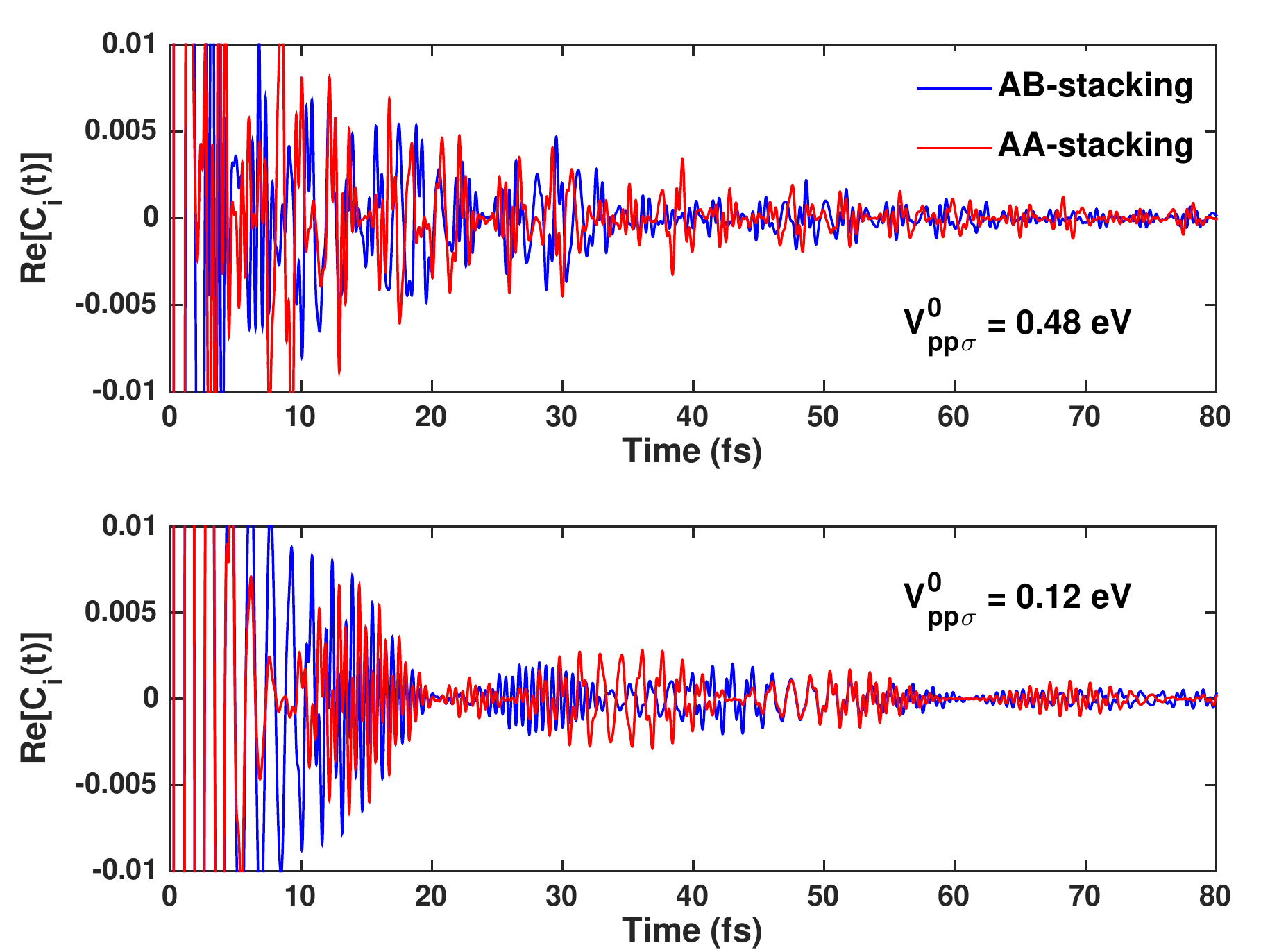}
\caption{\label{Fig6}Time auto-correlation functions $C_i(t)$ of the AA- (red) and AB-stacking (blue) bilayer graphene calculated using the NN model with $V^0_{pp\sigma} = 0.48$ eV and 0.12 eV.}
\end{center}
\end{figure}
%
%
We now analyse the features of the intra-layer spreading patterns in the AA- and AB-configurations. In Figs.~\ref{Fig4} and \ref{Fig5} we present the evolution of a $2p_z$ state initially localised at a lattice node in the top layer of the two AA- and AB-stacking configurations of bilayer graphene, respectively. We use colours to represent the probability densities on two graphene layers, specifically, the red for the top layer and the black for the bottom one. Similar to the case of monolayer graphene, the radius of the solid-circle at each lattice node is proportional to the value of $P_j$ normalised at the maximum value for each value of time.

By comparing the wavefront spreading behaviour of the electron wave function in the monolayer graphene and that in the AA-stacking configuration for the evolution time  $t < 1.3$~fs, we realise that the distribution of $P_j(t)$ on the top and bottom graphene layers are similar to the case of  monolayer graphene, but becomes different for larger evolution time. It should be noticed from Fig.~\ref{Fig3}(a) that in the duration of $(0, 1.3)$~fs the top layer-integrated probability density $\mathcal{P}_\text{T}$ monotonically decreases. It means that the wave continues transferring to the bottom layer and achieves the maximal transferring percentage at $1.3$~fs. When continuing to increase $t$, the part of the wave in the bottom layer transfer back to the top one. It results in the  oscillation behaviour of $\mathcal{P}_\text{T}(t)$ and $\mathcal{P}_\text{B}(t)$ similar to a Fabry-P\'erot resonator. From Fig.~\ref{Fig3}(a) we can determine a set of characteristic time intervals, \emph{e.g.}, $(0,1.3)$~fs, $(2.4,4.6)$~fs, $(6.1, 7.3)$~fs, and so on, in which the wave transfers predominantly from the top to the bottom layer; alternatively, in the complementary time intervals the wave transfers  in the opposite direction. We found that after $1.3$~fs,  the distribution of $P_j(t)$ of monolayer graphene is neither coincident to that on the top nor the bottom layers of the bilayer system. This difference is the result of the combination of the intra-layer and inter-layer spreading induced by the hopping terms in the Hamiltonian~\eqref{eq2}. 
The wavefront at long evolution time present a hexagonal shape similar to the monolayer graphene case. A direct inspection, however, shows that the form of the spreading pattern obeys only the point group $C_{3v}$. Remember that the symmetry of a node in the AA lattice is characterised by the point group $D_{3h}$ but, the successive interlayer penetration of the electron wave lowers the symmetry of the distribution of probability densities to the $C_{3v}$ symmetry.

For the AB-stacking configuration, we use the same technique for displaying data as for the AA configuration. From Fig.~\ref{Fig3}(b), we learn that for this configuration, the characteristic time $\tau_{\text{T}\to\text{B}}\approx7$~fs is larger than for the AA one. We found that when $t < 1.3$~fs, the probability densities on the top layer is in general  larger than those on the bottom one. 
The distribution of $P_j(t)$ on the top layer is identical to that of the monolayer graphene, but a quantitative difference becomes appearing when $t\in(1.3,2.1)$~fs. When $t > 2.1$~fs, the probability densities on the bottom layer become comparable to those on the top layer and different from those on the monolayer graphene in both quantitative and qualitative aspects. Referencing Fig.~\ref{Fig3}(b), the interval $(0, 1.3)$~fs is the one in which the wave monotonically transfers from the top layer to the bottom one. Though the percentage of the wave transfer at $t \approx 1.3$ is smaller than 50\%, successively the wave on the bottom layer transfers back to the top one. When this process occurs, it causes the change in the distribution of the probability densities from that of monolayer graphene.  From Fig.~\ref{Fig3}(b) we determine the sets of time intervals $(0,1.3)$~fs, $(2.2, 3.1)$~fs,  $(3.9, 6.6)$~fs, and so on, in which the wave transfers  predominantly from the top to the bottom layer, and in the complementary intervals where the wave transfers in the opposite direction. At long evolution time, the wave spreading is also characterised by a wavefront in the hexagonal shape that, similar to the AA lattice case, reflects the plane symmetries of the lattice nodes in the AB-stacking system, \emph{i.e.}, the group $C_{3v}$, a sub-group of the point group $D_{3d}$. 

We also calculated the time auto-correlation function $C_i(t)$ for the AA- and AB-stacking configurations. In Fig.~\ref{Fig6} we present the data for $C_i(t)$ as a function of the time-evolution for the two different parameter models:  $V^0_{pp\sigma}=0.48$ eV and $0.12$ eV. We observe, in general, the intricate behaviour of $C_i(t)$ in the two cases, which are different from each other. Interestingly, the figure shows the beating behaviour of the auto-correlation functions when considering $V^0_{pp\sigma} = 0.12$~eV. Our calculation shows that the beating behaviour does not appear clearly with $V^0_{pp\sigma} = 0.48$~eV, but it does when decreasing the value of $V^0_{pp\sigma}$ to the values smaller than about 0.3 eV. We also realise the beating oscillation behaviour is similar to the oscillation features of the time auto-correlation function of the monolayer graphene. It is expected because we should obtain a picture of the two independent graphene layers in the limit of $V^0_{pp\sigma}=0$. This observation reflects the fact that the interlayer coupling plays the role of modulating the electronic states between the two graphene layers. When a wave is spreading in one layer, it penetrates partially into the other and thus creates two waves spreading in the two layers. The coupling between the two layers induces the exchange of wave between the two layers and forms the wave-interference pattern in the space limited by the two layers.  The interference is sensitive to the alignment of the two atomic lattices. It thus explains the typical evolution features of electronic states in particular atomic configurations. Though the behaviour of the auto-correlation functions versus the time is complicated,  its Fourier transform results in the  density of states of these two configurations.~\cite{Do-2018,Weckbecker-2016}

\subsection{Twisted bilayer graphene}
Regarding the atomic structure, twisted bilayer graphene is a generalisation of the AA- and AB-stacking bilayer systems with a generic rotation angle between the two graphene layers. 
In general, the alignment between the two graphene lattice in the TBG systems is not commensurate, \emph{i.e.}, not defined by a unit cell, and thus the lattice has very low symmetry. In the case of commensurate stacking, the space group characterising the TBG lattice is determined to be either $p3m1$ or $p6mm$ depending on both the twist origin and the twist angles.~\cite{Zou-2018}. Interestingly, the generic TBG lattice shows a special moir\'e structure of the hexagonal form. In each moir\'e zone, we can find regions in which the atomic arrangement is close to the AA- and AB- or BA-stacking configurations. We illustrate the moir\'e zone in Fig.~\ref{Fig8} with the blue hexagon where we marked the AA- and the AB-like regions [c.f.~the frame with $t=0.2$~fs]. The AB-like regions are of two distinct types: one where the A sub-lattice is in the top layer and another one on with the B sub-lattice is in the top layer.
The AA-like and the two AB-like regions form two interpenetrating superlattices, a triangular and a honeycomb one, respectively.
We investigated the electron time-evolution in a series of TBG configurations with different twist angles.  The qualitative behaviour of the wave evolution is similar for the different twist angles we have investigated; thus we are going to present results for the case of two incommensurate twist angle $2.5^\circ$ and  $5^\circ$.
%
%
\begin{figure}
\begin{center}
\includegraphics[clip = true, trim = 1.8cm 7cm 2.5cm 7.2cm,width = \columnwidth]{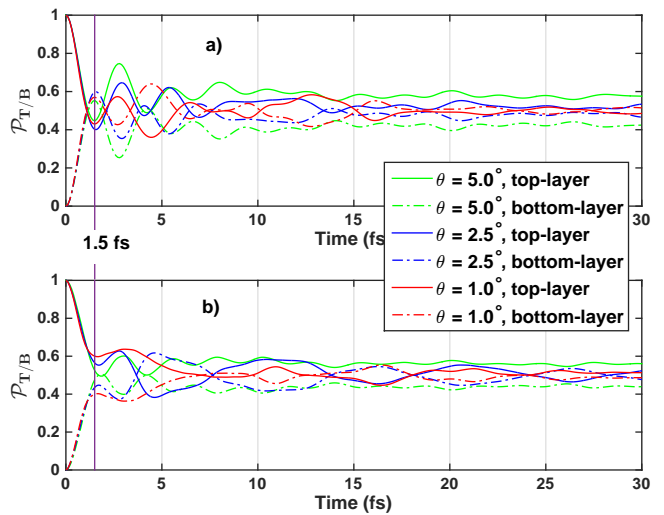}
\caption{\label{Fig7}Layer-integrated probability densities $\mathcal{P}_\text{T/B}(t)$ in the top (solid lines) and bottom (dot-dashed lines) layers of three TBG configurations with the twist angles of $5^\circ$ (green), $2.5^\circ$ (blue), and $1^\circ$ (red). Panels (a) and (b) are for the cases that the initial state $|2p_z\rangle$ locates at the central point of the AA$_0$-like and AB-like region, respectively. The parameter $V_{pp\sigma} = 0.48$ eV. The vertical lines highlight the time $t = 1.5$ fs discussed in text.}
\end{center}
\end{figure}
%
%
%
%
\begin{figure*}
\begin{center}
\includegraphics[width = \textwidth]{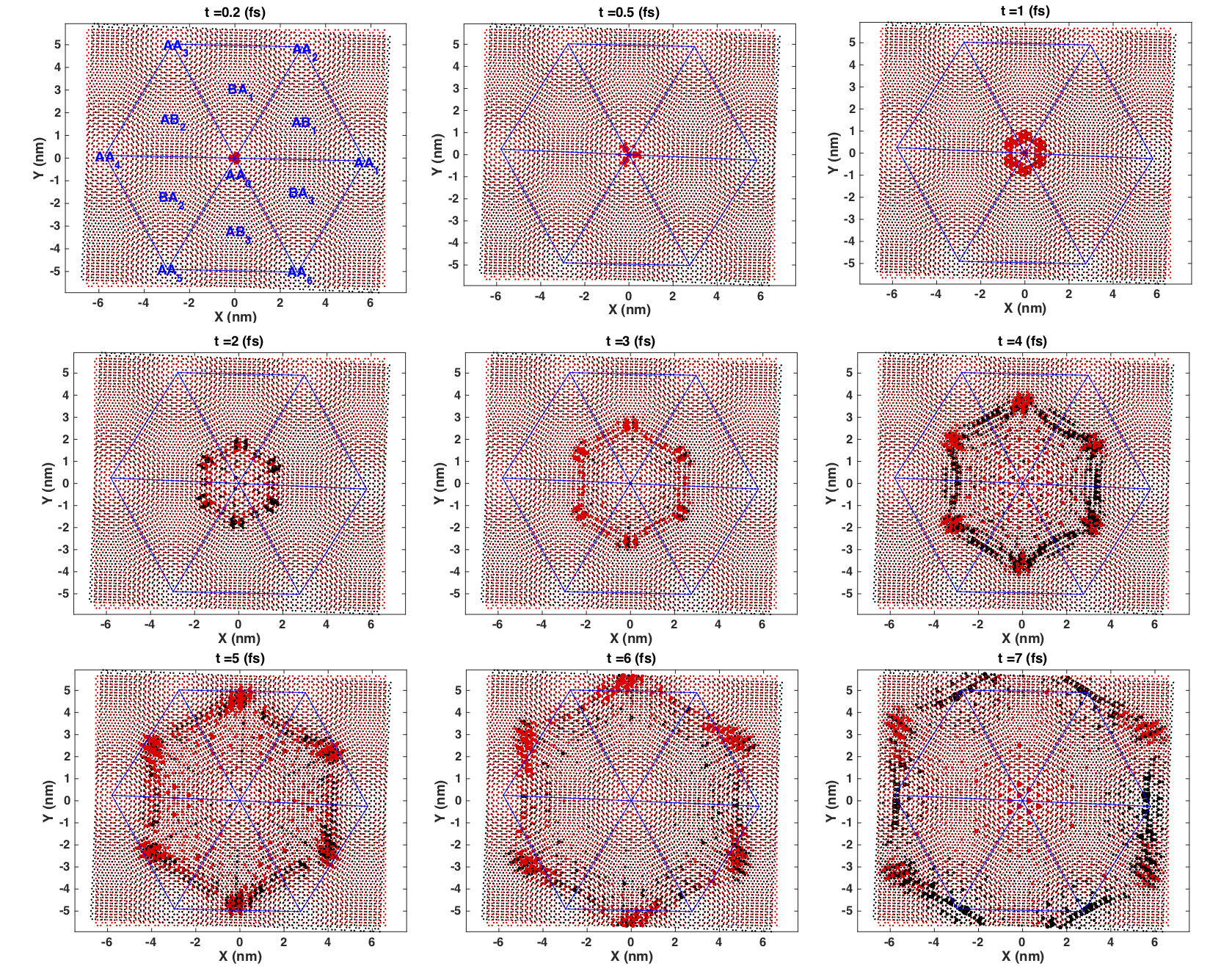}
\caption{\label{Fig8}Spreading of the electron probability density in the TBG configuration with $\theta = 2.5^\circ$ taken at several time moments. The initial state $|\psi(0)\rangle = |i\rangle$ located in the centre of the AA-like region in the moir\'e zone. Lattice nodes in red/black belong to the top/bottom layer. The blue hexagon denotes the moir\'e zone. The characters AA$_i$ ($i=1,...,6$), AB$_i$ and BA$_i$ ($i=1,2,3$) remark the AA- and AB-like regions in the moir\'e zone.}
\end{center}
\end{figure*}
%
%
\begin{figure*}
\begin{center}
\includegraphics[width =0.8\textwidth]{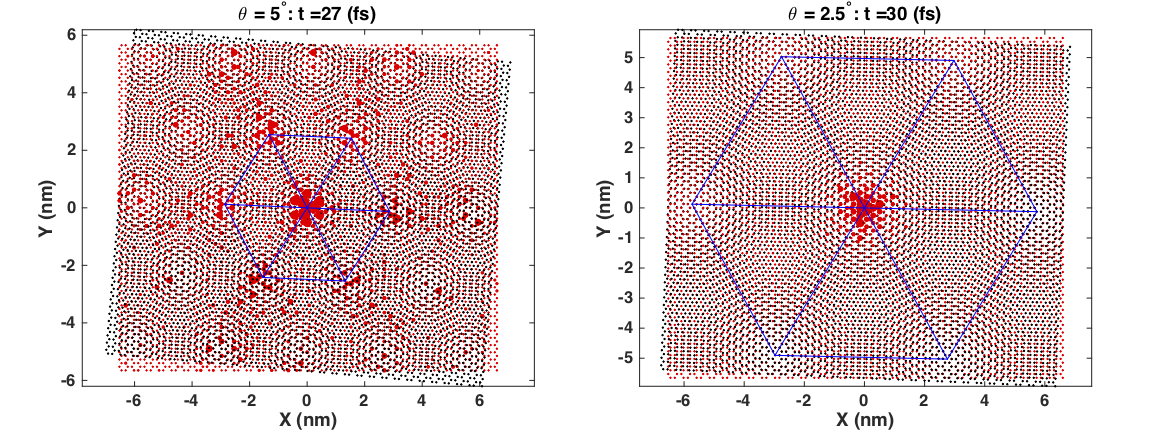}
\caption{\label{Fig9} Distribution of the probability densities in two TBG samples with $\theta = 5^\circ$ (left panel) and $\theta = 2.5^\circ$ (right panel) at large evolution times $t=27$~fs and $t=30$~fs.}
\end{center}
\end{figure*}
%
%
\begin{figure*}
\begin{center}
\includegraphics[width = \textwidth]{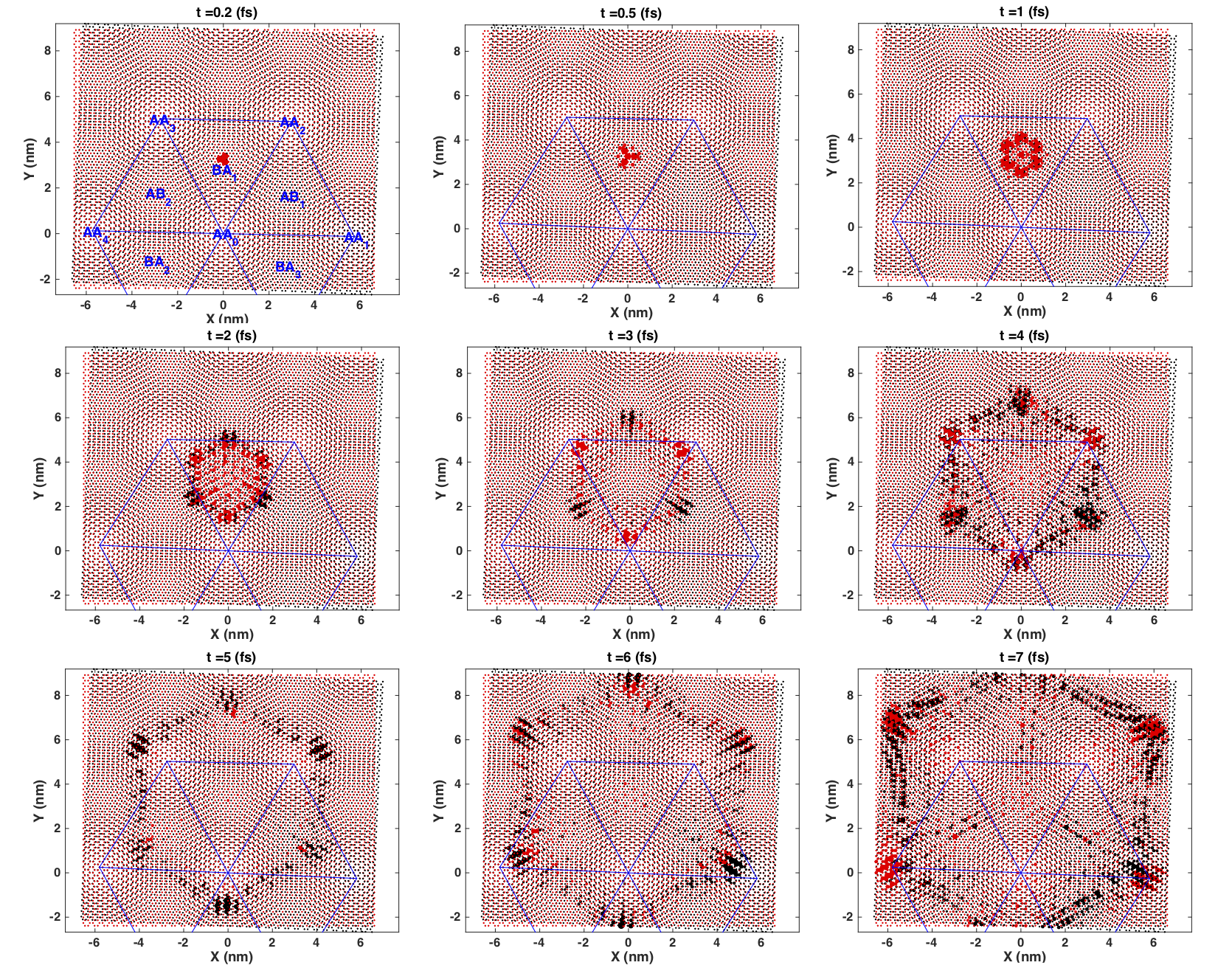}
\caption{\label{Fig10}Similar to Fig.~\ref{Fig9} but with the initial state $|\psi(0)\rangle = |i\rangle$ located in the centre of the AB-like region in the moir\'e zone. Lattice nodes in red/black belong to the top/bottom layer.}
\end{center}
\end{figure*}
%
%
%
\begin{figure}
\begin{center}
\includegraphics[clip = true, trim = 1.5cm 7cm 2cm 7cm,width = \columnwidth]{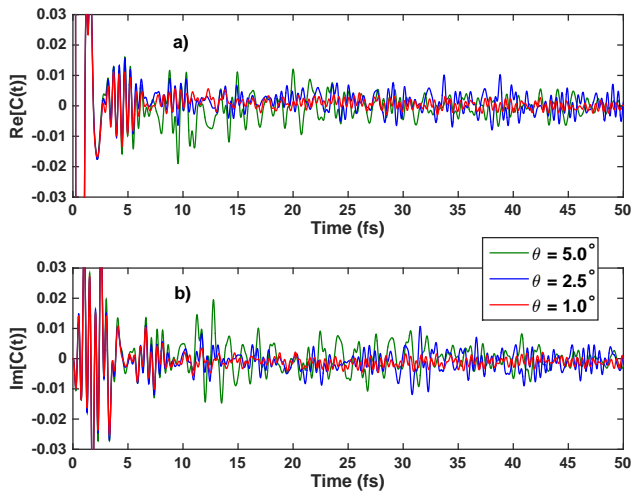}
\caption{\label{Fig11}The real (a) and imaginary (b) parts of the time auto-correlation function $C(t)$ for three TBG configurations with the twist angles of $\theta = 5^\circ$ (green), $2.5^\circ$ (blue) and $1^\circ$ (red).}
\end{center}
\end{figure}
%
%
The inter-layer coupling always induces the wave transfer between the two graphene layers. Figure~\ref{Fig7} shows that similar to the case of the AA-configuration, the transmission of electron wave function from the top layer to the bottom one reaches a maximal value in about 1.5~fs. To illustrate the wave transfer between two graphene layers, we study the variation of the layer-integrated probability densities on time. Figure~\ref{Fig7} shows the oscillation behaviour of the layer-integrated probability densities for three  incommensurate TBG configurations with the twist angle of $5^\circ$ (green), $2.5^\circ$ (blue) and $1^\circ$ (red). (The last one is close to the first magic angle $\theta \approx 1.05^\circ$.~\cite{Bristritzer-2011}) Furthermore, we investigated how the layer-integrated probability densities change by changing the initial position: the panels (a) and (b) are for the cases that the initial state localised in the centre of the AA$_0$-like and AB-like regions, respectively. 
We observe that the percentage of the wave transmitted from the top layer to the bottom one depends on the twist angle. For a short time of evolution, $t< 5$~fs the percentage is larger than 50\% in the configuration with the twist angle of $5^\circ$. However, after 5~fs, there is about 60\% of the wave propagating on the top layer and about 40\% doing in the bottom one. 
The minimal oscillation of the green curves $\mathcal{P}_\text{T/B}(t)$ implies a very weak transfer of wave between the two layers. This dynamical observation supplements to the explanation of the effective decoupling of the two graphene layers in the TBG configurations with large twist angles.\cite{Bristritzer-2011, Luican-2011, Brihuega-2012} In other words, the two parts of the wave become nearly independently propagating on the two graphene layers after a long time-evolution.  For the TBG configurations with much smaller twist angles, \emph{e.g.} $2.5^\circ$ and $1^\circ$, after about 10~fs, the fluctuation of the blue and red curves $\mathcal{P}_\text{T/B}(t)$ is always significant around the value of~50\%. It implies the strong interaction between the two wave components when propagating in the TBG lattices with small twist angles. 

We now present in Fig.~\ref{Fig8} the intra-layer spreading of a $2p_z$ state initially located at the central site of the AA-like region through the distribution of the probability densities $P_j(t)$. The time-evolution is illustrated similarly to the case of the AA- and AB-stacking configurations. From the figure we observe that during the time interval $(0,2)$~fs, the initial state spreads similarly to case of the AA-stacking lattice, \emph{i.e.}, extending along the six preferable directions,  then a hexagonal wavefront is established. In this time interval, the wavefront takes the typical hexagonal shape, and it is still within the AA-like region. 
For time larger than $t=3$~fs, we observe how the wavefront corners enter the AB-like regions, and the wavefront edges reach the transition regions between the AA$_j$ and AA$_0$ regions. At this time, the probability densities start to be redistributed: they become concentrated in the AB-like regions as the six clusters seen in the time frame at $t= 4$~fs.
These clusters then move in the transition regions between the AA$_i$ and AA$_{i+1}$ regions, \emph{i.e.}, along the zigzag lines connecting the AB-like regions in the first moir\'e zone to the other AB-like regions in the next moir\'e zones, while the probability densities on the edges of the hexagonal wavefront become scattered into the AA$_{i+1}$ ($i\ne 0$) regions, see the time frames at $t=5, 6$ and $7$~fs. Though the wavefronts on the two graphene layers take the same hexagonal pattern, the distribution of the probability densities on those does not obey the hexagon symmetry group. 
By inspection, we observe  the symmetry of the wavefront reduces to the ``approximate'' $C_3$ symmetry. The wavefronts on the two graphene layers are not coincident due to the misalignment of the lattice stacking. 

Interestingly, for long evolution time,  we observe the higher intensity of the probability densities in the AA-like regions ($t > 15$~fs), particularly, in the TBGs of tiny twist angles, this higher intensity is observed significantly in the AA$_i$ region only, see Fig.~\ref{Fig9}. This observation might reflect the ``localisation'' of low-energy Bloch wave functions in the AA-like regions as depicted in Refs.~[\onlinecite{Trambly-2010, Bristritzer-2011,Trambly-2012, Do-2018,Weckbecker-2016}]. Notice that, at the evolution time $t$, we would expect the dominance of electron states of energies about $\hbar/t$. It therefore implies that, at long observation time, the localised signature shown in Fig.~\ref{Fig9} of the electron wave function in the AA-like regions is the behaviour of the states associated with the narrow energy band around  the charge neutrality level. This localisation feature might also be related to topological properties of the wavefront as recently pointed out in Ref.~[\onlinecite{Kang-2018,Po-2018,Koshino-2018}]. Quantitatively, this association is consolidated by Fourier transforming the time auto-correlation $C(t)$ defined by Eq. (\ref{eq11}) to obtain the densisty of states. The resulted DOS of the TBG configurations with the twist angles $\theta < 2.5^\circ$ shows a small, but significant peak, around the charge neutrality level as reported in Refs.~[\onlinecite{Do-2018,Weckbecker-2016}].

In the following, once again, we will show how the dynamics of wave spreading in TBG strongly depends on the symmetry of an initial localised state. We now consider an initial $2p_z$ state at the central node of one AB-like region in the moir\'e zone. We note that, contrary to the central node in the AA-like regions, for this choice, there is no exact symmetry elements containing the central node of the AB-like regions. 
For short time-evolution ($t<1$~fs), the wave spreading is similar to that in the AB-stacking lattice. When increasing the evolution time the wave evolves preferably in the directions heading three next-neighbour AB-like regions, \emph{i.e.}, along the zigzag lines separating the AA-like regions (c.f.~panel $t=2$~fs in Fig.~\ref{Fig10}). Along the opposite directions, the wave spreads into the AA-like regions, and the probability densities become concentrated at the centre rather than scattered, (c.f.~panels for $t=3$ and $4$~fs in Fig.~\ref{Fig10}). 
Following the distribution of the probability densities at larger times, the probability densities propagate along the zigzag lines in the transition regions between the AA$_i$- and AA$_{i+1}$-like regions and concentrated in the centre of the AA-like regions. Due to the ``approximated'' symmetries about the initial position of the $2p_z$ state, the wavefront is formed and has an almost hexagonal shape. The six corners of the wavefront orient the preferably evolved directions.
The distribution of $P_j(t)$ on the two layers satisfies the ``approximated"  point group symmetry $C_3$.

To complete our discussion of the wave evolution in the TBG lattices, we present in Fig.~\ref{Fig11} the time auto-correlation function $C(t)$.  We remind again that the choice of the initial condition affects crucially the wave spreading. In  Figs.~\ref{Fig8} and~\ref{Fig10}, we have presented the data for two particular initial conditions which result in typical spreading patterns of the $2p_z$ state. 
In order to extract quantitative information on observables, for instance the density of states, from the time-evolution of electronic states we have to account for all possible initial conditions. According to Eq.~\eqref{eq11}, to calculate the time auto-correlation function $C(t)$, we need to calculate a set of functions $C_i(t) = \langle i|\psi(t)\rangle$ with the initial states $|i\rangle = |2p_z\rangle$ chosen at every lattice node in a sufficiently large TBG sample. 
Though a TBG lattice is not always defined by a unit cell with translational symmetry, the moir\'e zone can be seen as an approximated unit cell.  It suggests that we need to consider only the lattice nodes in a moir\'e zone. However, since the typical length $L_\text{M}$ defining the size of the moir\'e zone is related to the twist angle $\theta$ via the expression $L_\text{M} = \sqrt{3}a_\text{cc}/2\sin(\theta/2)$, it means that we have to work with a very large moir\'e for the TBG configurations in the case of tiny twist angle | this can be a difficult task in practice. However, we demonstrated in Ref.~[\onlinecite{Do-2018}] that an appropriate sampling scheme for a moderate number of lattice nodes in the moir\'e zone is sufficient to obtain reliable values for important physical observables. We apply here the same scheme to evaluate $C(t)$. The results are shown in Fig.~\ref{Fig11} for three TBG configurations with $\theta = 5^\circ, 2.5^\circ$ and $1^\circ$. The figure shows the complex behaviour of $C(t)$ as a function of time. Despite that, the Fourier transform of $C(t)$, see Eq.~\eqref{eq12}, results in the density of states with typical van Hove peaks are shown in Ref.~[\onlinecite{Do-2018,Weckbecker-2016}].

\section{Conclusion}\label{conclusions}
We have presented a study of the time-evolution characteristics of electrons in the bilayer graphene lattices with arbitrary twist angles. We used the Chebyshev polynomials of the first kind to approximate the time-evolution operator for a sufficiently long time-evolution to calculate time-correlation functions reliably. We have shown that the inter-layer electronic coupling induces the interchange transfer of waves between the two graphene layers, resulting in the oscillating behaviour of the layer-integrated probability densities as a function of time, similar to complex Fabry-P\'erot oscillations. 
This behaviour can be also interpreted as the precession of electrons when describing the moir\'e-induced spatial modulation in the interlayer coupling in terms of non-Abelian gauge fields.~\cite{Jose-2012}
The percentage of the wave transmitted from one layer to the other depends on the twist angle, \emph{i.e.}, smaller than 50\% and weak oscillation for large twist angles, $\theta > 2.5^\circ$,  and larger than 50\% and strong oscillation otherwise. This dynamical observation supplements the understanding of the effective decoupling between the two graphene layers in the TBG configurations with large twist angles. For the wave spreading in each graphene layer, we have indicated that the spreading shape of electron waves is dictated by the dominant hopping mechanism of the honeycomb pattern of the monolayer lattice and by the plane symmetries of the bilayer lattices. 
The wave spreading is irregular and takes place in two stages: The first one occurs within a very short time-evolution, in which the wave spreads to the three nearest neighbours and then develop to the lattice nodes along the directions of the armchair-lines of the honeycomb lattice. The second stage is characterised by the formation of a well-defined wavefront of hexagonal shape with the corners developing faster the edges. 
For tiny twist TBG configurations, we have observed the signature of the electron localisation in the AA-like regions inside the TBG's moir\'e zone at long time-evolution. This would associate with the formation of a narrow energy band around  the charge neutrality level. We have shown the interchange transfer of wave between the two graphene layers resulting in the difference of the distribution of the probability densities on the TBG lattices from that on the monolayer. We have also observed the appearance of a beating pattern in the autocorrelation functions for a reduced intra-layer coupling | it is possible to achieve this reduction experimentally.~\cite{Jeon-2018} It might suggest a way for engineering the electronic properties of the bilayer systems. This study provides a complementary intuitive understand of the electron behaviours in the twisted bilayer graphene. The calculation method implemented here represents an alternative paradigm for future studies of exotic electronic properties of layered materials, including twisted bilayer graphene but also other van der Waals heterostructures.

\section*{Acknowledgements}
The work of VND and HAL is supported by the National Foundation for Science and Technology Development (NAFOSTED) under Project No. 103.01-2016.62. The work of DB is supported by Spanish Ministerio de Ciencia, Innovation y Universidades (MICINN) under the project FIS2017-82804-P, and by the Transnational Common Laboratory \textit{Quantum-ChemPhys}.

\bibliography{bibliography}

\end{document}